\documentclass{llncs}
\usepackage{makeidx}  % allows for indexgeneration
\usepackage{cite}
\usepackage{xspace} 
\usepackage{amsmath,amssymb,amsfonts}
\usepackage{algpseudocode}
\usepackage{graphicx}
\usepackage{multirow}
\usepackage{textcomp}

\usepackage[table]{xcolor}

\newcommand{\cb}[1]{\cellcolor{blue!25} #1}
\newcommand{\crd}[1]{\cellcolor{red!25} #1}

\usepackage{listings}

\definecolor{keywordcolor}{rgb}{0.13,0.13,1}
\definecolor{greenComments}{RGB}{63,127,95}
\lstdefinelanguage{PSEUDO}{
    keywordstyle=[2]\color{blue},
    keywords=[2]{algorithm, is, input, output, for, each, in, do, while, return, if, then, else, end, and, or},
    sensitive=false,
    morestring=[b]",
    morecomment=[l]{//},
    }

\lstdefinestyle{pseudo-style}{
    language=PSEUDO,
    backgroundcolor=\color{white},   
    commentstyle=\color{purple},
    keywordstyle=\color{blue},
    basicstyle=\ttfamily\scriptsize,
    breakatwhitespace=false,         
    breaklines=true,                 
    captionpos=b,                    
    keepspaces=true,                 
    showspaces=false,                
    showstringspaces=false,
    showtabs=false,                  
    tabsize=2,
    columns=fullflexible,
}

\lstnewenvironment{pseudo-listing}[1][]{
    \lstset{style=pseudo-style, caption=#1, mathescape=true }
  }
  {}
  
\begin{document}

\def \midd {COPERNIC}
\def \middleware {\emph{\midd}\xspace}
\mainmatter              % start of the contributions
\title{Cognitively-inspired Agent-based Service Composition for Mobile \& Pervasive Computing}
\titlerunning{Cognitively-inspired Service Composition for Mobile Computing}  % abbreviated title (for running head)
%                                     also used for the TOC unless
%                                     \toctitle is used
%
\author{Oscar J. Romero\inst{1}\orcidID{0000-0002-4470-8209}}
\authorrunning{Oscar J. Romero} % abbreviated author list (for running head)
\institute{Carnegie Mellon University, Pittsburgh OA 15213, USA,\\
\email{oscarr@andrew.cmu.edu}, \texttt{http://www.cs.cmu.edu/~oscarr/}
}

\maketitle              % typeset the title of the contribution

\begin{abstract}
Automatic service composition in mobile and pervasive computing faces many challenges due to the complex and highly dynamic nature of the environment. Common approaches consider service composition as a decision problem whose solution is usually addressed from optimization perspectives which are not feasible in practice due to the intractability of the problem, limited computational resources of smart devices, service host's mobility, and time constraints to tailor composition plans. Thus, our main contribution is the development of a cognitively-inspired agent-based service composition model focused on bounded rationality rather than optimality, which allows the system to compensate for limited resources by selectively filtering out continuous streams of data. Our approach exhibits features such as distributedness, modularity, emergent global functionality, and robustness, which endow it with capabilities to perform decentralized service composition by orchestrating manifold service providers and conflicting goals from multiple users. The evaluation of our approach shows promising results when compared against state-of-the-art service composition models.
\end{abstract}
%
%% ================== INTRODUCTION ====================== %%
\section{Introduction and Motivation}
\label{sec_intro}
In Mobile and Pervasive Computing (MPC), a \emph{Service} can be defined as any hardware or software functionality (resources, data or computation) of a smart device that can be requested by other devices for usage~\cite{[6]Raychoudhury:2013}. \emph{Service composition} refers to the technique of creating composite services by the aggregation of atomic, simpler and easily executable services. %Research in Service Composition has predominantly followed two directions~\cite{Chakraborty:2005}. One direction defines languages to formally specify services and composite services (such as BPEL4WS, OWL-S, and WSDM) in terms of service input/output, service pre and postconditions, fault handling, service invocation mechanisms, and development of complex planning engines~\cite{Sirin:2004,Tari:2010}. The other direction of research aims at developing middleware~\cite{Ylianttila:2012,Karunamurthy:2012,Paik:2014,Hadj:2017} that enable service composition. These middleware assume a declarative specification of a composite service and perform the task of discovering, integrating and executing the actual services. Our paper focuses on the latter direction. 
Despite the existence of MPC middleware for automatic service composition~\cite{[3]Ibrahim:2009,[6]Raychoudhury:2013,[4]Stavropoulos:2013,[5]Immonen:2014}, there are still some challenges that need to be tackled, as we illustrate in the next example: 

\emph{Alice and Bob are planning to have a theme party at their home next weekend (high-level goal), so they need to coordinate some tasks among them. To achieve this goal, they interact with a user application (e.g., a personal assistant~\cite{Tomazini:2017}, a chatbot~\cite{Zhao:2018}, etc.) connected to a MPC middleware installed on their mobile and wearable devices (e.g., smartphones, smartwatches, tablets, etc.), which act as Service Providers. The high-level goal, which will lead to the creation of a composite service, may be decomposed into 3 sub-goals: buy-food, buy-beer, and buy-home-decoration. There is also a set of atomic services that are hosted by service providers: get-location, find-place, calculate-distance, who-is-nearer, share-shopping-list, go-to-place. Now, each sub-goal is accomplished by the composition of a sequence of services, e.g.: buy-food = \{get-location(user) $\rightarrow$ find-place(supermarket) $\rightarrow$ calculate-distance(user, supermarket) $\rightarrow$ who-is-nearer( market, users) $\rightarrow$ share-shopping-list(users) $\rightarrow$ go-to-place(user, market)\}.}

Given the previous example we focus on five challenges, so service composition should:
(1) consider preferences from multiple users; (2) coordinate the interaction between services hosted by different service providers; (3) consider resource scarcity in smart devices~\cite{Romero:2018};
(4) perform dynamic adaptation to unpredictable changes occurring in the environment; (5) deal with both short-term and long-term user's goals.
Performing service composition while taking into account a myriad of variable factors as described above (e.g., users, services, service providers, QoS values, context, etc.), makes the problem become intractable even for approaches that use dynamic composition. The main issue with these approaches is that they propose solutions focused on optimality (e.g, graph-, rule-, and workflow-based solutions), which do no consider limitations imposed by the decision-making process (specially on smart devices), revealing their inability to process and compute the expected utility of every alternative action when variable factors grow in size (combinatorial explosion). Therefore, we propose a cognitively-inspired approach based on bounded rationality, which centers on the fact that perfectly rational decisions are often not feasible in practice because of the intractability of the problem, the limited computational resources, and time constraints; instead, our approach seeks satisfactory solutions rather than optimal ones. 
Thus, our main contributions are twofold: (1) We propose \middleware, a cognitively-inspired agent-based service composition middleware, as a first approach to addressing the five challenges described above (i.e., multiple users, decentralized coordination, and inexpensive, dynamic, and long-short term composition) using a bounded rationality approach; and (2) We develop a prototype of \middleware and evaluate its performance against to state-of-the-art service composition models. 
This paper is organized as follows:  Section~\ref{sec_service_model} outlines the system architecture. Section~\ref{sec_middleware} details our approach, and Section~\ref{sec_evaluation} reports the experimental results. Sections~\ref{sec_related_work} and~\ref{sec_conclusions} presents the related work and the conclusions, respectively.

%% ================== Overview ====================== %%
\vspace{-0.4cm}
\section{Overview}
\label{sec_service_model}
\vspace{-0.1cm}

%% ===================== Preliminaries ====================== %%
\vspace{-0.2cm}
\subsection{Preliminaries}
\label{sec_preliminaries}
\vspace{-0.2cm}

As considered in the literature~\cite{Balzer:2004}, we distinguish two types of services: \emph{abstract} and \emph{concrete} services. Formally, a concrete service $cs_{i}$ is a tuple  $\langle cs_{i}^{in}, cs_{i}^{out},$ $cs_{i}^{prec}, cs_{i}^{postc}, cs_{i}^{QoS}, cs_{i}^{ctx} \rangle$ that performs a functionality by acting on input data ($cs_{i}^{in}$) to produce output data ($cs_{i}^{out}$), with pre-conditions ($cs_{i}^{prec}$), post-conditions ($cs_{i}^{postc}$), Quality of Service requirements ($cs_{i}^{QoS}$), and contextual information. An abstract service $as_i$ is a tuple $\langle as_{i}^{pre}, as_i^{post}, as_i^{cs} \rangle$ realized by several concrete services that offer the same functionality ($as_i^{cs} \in \{cs_{(i,1)}, cs_{(i,2)},...,$ $cs_{(i,n)}\}$)  with preconditions ($as_{i}^{pre}$) and postconditions ($as_{i}^{post}$) such that $\forall cs_{(i,j)},$ $cs_{(i,k)} \in as_{i}^{cs} / (as_{i}^{pre} = cs_{(i,j)}^{pre} \cap cs_{(i,k)}^{pre}) \wedge (as_{i}^{post} = cs_{(i,j)}^{post} \cap cs_{(i,k)}^{post})$.

\vspace{-0.3cm}
\subsection{System Architecture}
\vspace{-0.2cm}
\label{sec_scm}

\begin{figure}[tbp]
\centerline{\includegraphics[width=\textwidth]{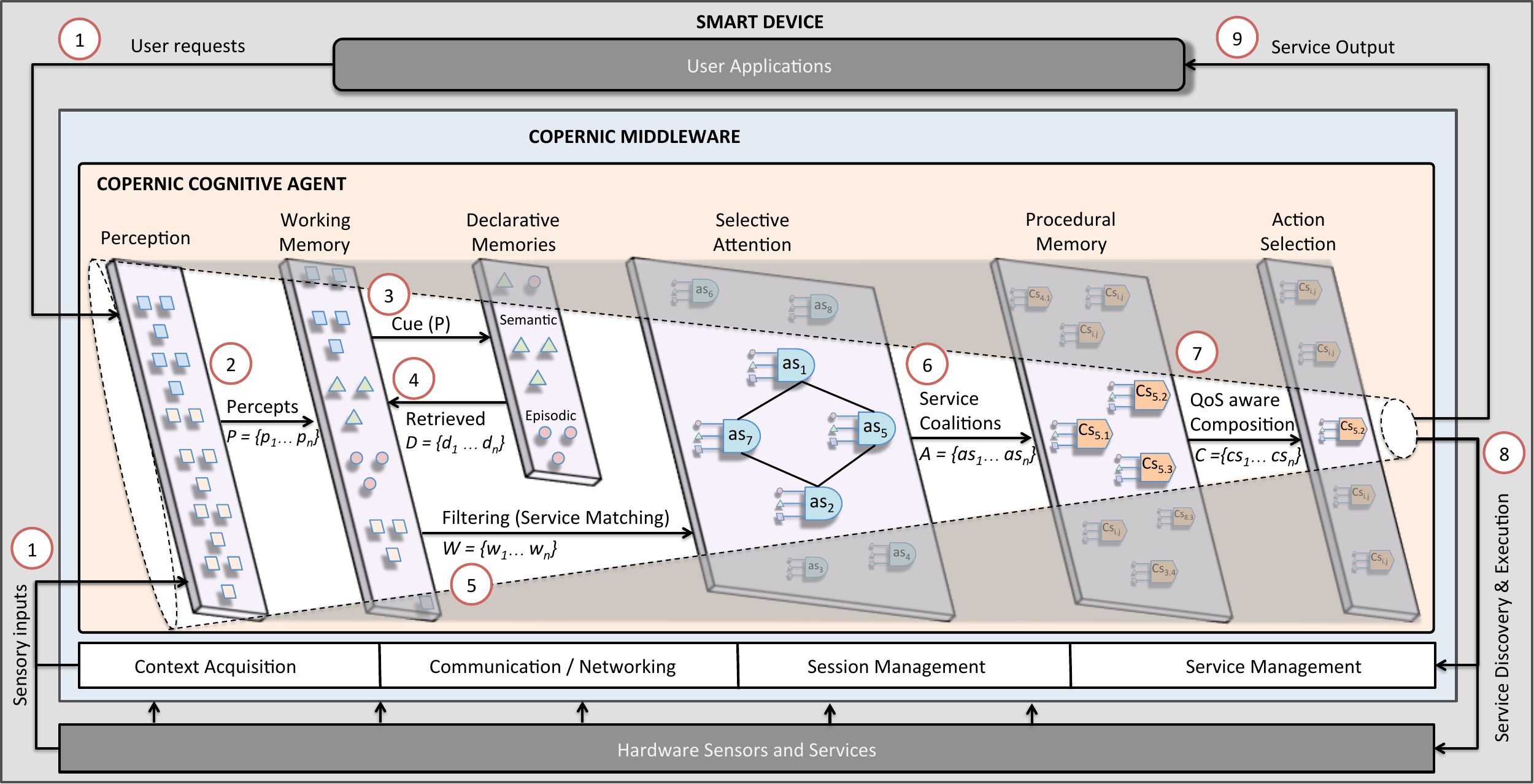}}
\vspace{-0.2cm}
\caption{\middleware's overall Architecture (Single Device). The white cone illustrates how a continuous stream of data is filtered out so only the most relevant elements are retained for the composition while the others are either discarded or put on hold until they receive more activation to become participants.}
\vspace{-0.5cm}
\label{fig_architecture}
\end{figure}

Figure~\ref{fig_architecture} depicts the overall architecture of \middleware, though it is worth noting that it does not reflect yet the distributed nature of our model. The \middleware Agent is a cognitive module inspired by architectural principles defined by the Common Model of Cognition (CMC)\cite{llc:2017,Romero:ADL:2018}, a computational model that captures a consensus about the structures and processes that are similar to those found in human cognition. Next, we briefly describe \middleware's pipeline (Figure~\ref{fig_architecture}) and its realization on the CMC model.
In \emph{Step 1}, the \emph{Perception} module makes sense of the agent's current state by processing both external (e.g., user requests) and internal (i.e., signals from other modules) sensory inputs, categorizing that information, and recognizing situations where a set of abstract services may be triggered. 
In \emph{Step 2}, the \emph{Perception} module outputs a set of symbolic structures (percepts) that are stored in a Working Memory (WM) for further processing as abstract service inputs. 
In \emph{Step 3}, the WM cues the declarative memories (i.e., \emph{Episodic Memory} that retrieves information about historic performance of services, context, etc., and \emph{Semantic Memory} that retrieves service definitions, user preferences, etc.) and stores local associations in \emph{Step 4}. 
In \emph{Step 5}, the content of the WM is filtered out by the attention mechanism so the agent only focuses on the most relevant information needed for matching abstract services. 
In \emph{Step 6}, goals are decomposed and abstract services compete and cooperate (creating coalitions) among them in order to get the focus of attention.
In \emph{Step 7}, the \emph{Procedural Memory} executes a set of heuristics to dynamically bind abstract services to concrete services by validating the QoS requirements.
In \emph{Step 8}, the \emph{Action Selection} chooses the most appropriate action to execute a concrete service using discovery protocols adapted to the heterogeneous nature of the environment (the process is repeated until all sub-goals are satisfied). 
In \emph{Step 9}, service's output is returned to the application. 
Unlike traditional approaches that create uPFRont composition plans which are prone to inadaptability, in our approach, plans emerge from the interaction of cascading sequences of \textbf{cognitive cycles} corresponding to perception-action loops (steps 1-8) where compositional conditions are validated in every cycle. This contribution allows service composition to be reactive, robust, and adaptive to dynamic changes while composition plans are generated on-the-fly by using minimal resources as a result of filtering out a continuous stream of data.

%% ===================== COPERNIC ====================== %%
\vspace{-0.45cm}
\section{Approach}
\label{sec_middleware}
\vspace{-0.3cm}

%% ===================== Perception ====================== %%
% \subsection{Perception}
% \label{sec_perception}

\vspace{0.1cm}
\noindent\emph{A. Perception}

The perception module defines a set of feature detectors in charge of detecting and classifying the sensory inputs, i.e., there are different feature detectors to identify external stimuli (i.e., user requests) and internal stimuli (i.e., user's context, physical context -- sensor readings, and service QoS). Feature detectors create \emph{percepts}, units of perceived information with a symbolic representation (key-value pairs, e.g., \texttt{<location, home>}) and an activation level. Percepts activation rapidly decay over time (when not re-stimulated) according to the following inverse sigmoid function: $p_{act_{i}} = \frac{sal_{i}}{\log_{2} cc_{i}}$, where $p_{act_{i}}$ is the activation for percept $i$, $sal_{i}$ is the salience of the stimulus (the quality by which a percept stands out from its neighbors), and $cc_{i}$ the number of cognitive cycles since the last time the percept received activation. Salience is a numeric value between 1 - 10 (being 10 the most salient stimulus) and serves for designers to add some relevance to perceived information, e.g, a ``user request'' percept may have a higher salience (so it should last longer) whereas ``Temporary WiFi disconnection'' may have a lower salience. This module outputs a set of percepts $P = \{p_{1}...p_{n}\}$ $|$ $\forall p_{i} \in PR$, where $PR$ is a set of premises such that $as^{pre} \cup cs^{pre} \subseteq PR$.

%% ===================== WM ====================== %%
% \subsection{Short-term Working Memory (WM)}
% \label{sec_wm}

\vspace{0.1cm}
\noindent\emph{B. Short-term Working Memory (WM)}

WM holds previous percepts not yet decayed away, and local associations from declarative memories that are combined with the percepts to understand the current state of the composition. Information written in the WM may reappear in different cognitive cycles until it fades away. To that purpose, WM defines a limited storage capacity (default: 7 units~\cite{Miller:56}) and a recency-based decay function that keeps active a limited number of units, expressed as a base-level activation function\cite{Anderson:2004}: $B_{i}^{w} = iB_{i}^{w} + \sum_{l = 1}^{n} t_{l}^{-d}$, where $i$ is a WM unit ($w_{i}$), $l$ is the $l$th setting of $w_{i}$, $t_{l}$ is the time since $l$th unit was presented, $iB_{i}^{w}$ is the initial value of activation, and $d$ is a decay parameter that reflects differences in WM units volatility, e.g., dynamic changes on \emph{user context} happen more often than changes on \emph{user preferences}, so the former should have a higher decay value, whereas the latter should have a lower one, which makes user context obsolete quicker than user preferences. If $B_{i}^{w}$ is above a threshold ($B_{i}^{w} > t_{w}$) then $w_{i}$ will be used as an input for service matching, i.e., $w_{i} \subseteq (as^{pre} \cup cs^{pre}) \subseteq PR$.

%% ===================== Episodic ====================== %%
% \vspace{-0.26cm}
% \subsection{Long-term Episodic Memory (EM)}
% \label{sec_episodic}
% \vspace{-0.15cm}

\vspace{0.1cm}
\noindent\emph{C. Long-term Episodic Memory (EM)}

EM is a content-address-able associative memory that records temporal sequences of user and system events. \middleware defines 3 types of EM: (1) \emph{User EM} that stores user past actions (e.g., Bob searched for nearby beer shops after doing the shopping); (2) \emph{Service Provide EM} that stores historic data of service and service provider performance (efficiency, reliability, QoS, reputation, failures, etc.); and (3) \emph{Network EM} that stores neighboring updates, network hops, etc. Any unit written to the WM cues a retrieval from EM, returning prior activity associated with the current entry. We used a Sparse Distributed Memory (SDM)~\cite{Kanerva:1988}, a high-dimensional space that mimics a human neural network. SDM is lightweight, random (it retrieves service associations in equal time from any location), content-addressable (it finds complete contents using content portions), and associative (it finds contents similar to a cue).

%% ===================== Episodic ====================== %%
% \subsection{Long-term Semantic Memory (SM)}
% \label{sec_semantic}

\vspace{0.1cm}
\noindent\emph{D. Long-term Semantic Memory (SM)}

SM is intended to capture the meaning of concepts and facts about: (1) service descriptions; (2) the world (e.g., \texttt{<Home><is-a><place>}); and (3) the user (e.g., preferences, goals, etc.). SM is implemented using a Slipnet, an activation passing semantic network, where each concept is represented by a node, and each conceptual relationship by a link having a numerical length, representing the ``conceptual distance'' between the two nodes involved, which is adjusted dynamically. The shorter the distance between two concepts is, the more easily pressures can induce a slippage (connection) between them. Nodes acquire varying levels of activation (i.e., measure of relevance to the current situation in the WM) and spread varying amounts of activation to neighbors. 
Let $B_{i}^{s} = B_{i-1}^{s} + \sum_{j=0}^{n} (k - L_{i,j})$ such that $B_{i}^{s}$ and $B_{i-1}^{s}$ are the current and previous activations of node $i$, respectively; $k$ is a constant for regulation of spreading activation; and $L_{i,j}$ is the conceptual length between nodes $i$ and $j$. 
Traditional semantic-driven approaches for service composition use ontology-based description languages, however, these static representations do not account for the dynamicity of the environment, require the use of semantic reasoners and ontologies, and lack a mechanism to represent conceptual distance. 
The declarative module outputs a set of premises $D = \{d_{1}...d_{n}\}$ $|$ $\forall d_{i} \in PR$.

% \begin{figure}[tb]
% \centering
% \includegraphics[width=8cm] 
% {images/semantic}
% \vspace{-0.2cm}
% \caption{Semantic Memory example for \emph{buy-beer} service. Since Bob likes imported beers and he is of legal age, \middleware is most likely to recommend Bob go buying the beers instead of Alice. \emph{do-shopping} service is farther (in terms of conceptual distance) than \emph{buy-beer} service.}
% \vspace{-0.55cm}
% \label{fig_semantic}
% \end{figure}

%% ===================== Attention ====================== %%
% \vspace{-0.1cm}
% \subsection{Selective Attention (SA)}
% \label{sec_attention}
% \vspace{-0cm}

\vspace{0.1cm}
\noindent\emph{E. Selective Attention (SA)}

% \begin{figure}[tb]
% \centering
% \includegraphics[width=8cm] 
% {images/attention}
% \vspace{-0.2cm}
% \caption{Selective Attention (SA). It depicts a portion of a BN with 6 services (s1...s6) and 1 goal. WM units $w_{1}$ and $w_{2}$ are preconditions of s1. $w_{3}$ is a positive postcondition (add list) of s1 and a precondition of s2, so both successor and predecessor links are created for s1 ad s2. An inhibitory link for $w_{2}$ avoids s1 to reactivate indefinitely (delete list). After \middleware suggests Bob go to the beer shop, he starts walking, so $w_{6}$ is set as a negative postcondition of s4 that in turn inhibits s5 (due to Bob can only go to a place at a time). Only s1, s2 and s3 have accumulated enough activation to be under the focus of attention. Due to spreading activation dynamics, the SA will activate this service sequence during the composition process: $s1 \rightarrow s2 \rightarrow s3$.}
% \vspace{-0.55cm}
% \label{fig_attention}
% \end{figure}

Based on Posner's theory of attention~\cite{Posner:2011}, our SA filters out a continuous stream of content from WM while carrying out three attentional functions: (1) maintaining an alert state (e.g., SA gives priority to salient information like context and QoS); (2) focusing agent's senses on the required information (e.g., to discover \emph{get-location} service and bring it into composition, SA needs to focus on changes on GPS readings); and (3) the ability to manage attention towards goals and planning (e.g., SA focuses on the high-level goal ``plan a party at home'' and its corresponding sub-goals).
SA uses a Behavior Network (BN)~\cite{maes}%,Romero:2011
, a hybrid system that integrates both a connectionist computational model and a symbolic, structured representation. %
%From connectionism it inherits properties of intrinsic parallelism, fault-tolerance, sophisticated retrieval and matching capabilities, density and global emergent computation from uniform local interaction rules. From symbolic AI, it adopts representation and structuring principles while avoiding problems of traditional AI solutions such as seriality/slowness, brittleness, rigidity, and the communication complexity of distributed AI systems.
%
BN defines a collection of behaviors (nodes) that compete and cooperate among them (through spreading activation dynamics) in order to get the focus of attention. In \middleware, each behavior maps to a single abstract service $as$, and \emph{``service discovering/matching''} is modeled as an emergent property of activation$/$inhibition dynamics among all abstract services. 
Revisiting the formal definition of an abstract service $as_i$, we have the tuple: $\langle as_{i}^{pre}, as_{i}^{add}, as_{i}^{del}, as_{i}^{\alpha} \rangle$, where $as_{i}^{pre}$ is a list of preconditions that have to be true before the service becomes active, $as_{i}^{add}$ and $as_{i}^{del}$ represent the expected (positive and negative) postconditions of the service in terms of an ``add'' and a ``delete'' lists, and $as_{i}^{\alpha}$ is the level of activation. 
If a WM unit $w_{i}$ is in $as_{i}^{pre}$ then there is an active link from $w_{i}$ to $as_i$. If the goal $g$ (i.e., a user request or any sub-goal stored in WM) is in $as_{i}^{add}$ then there is an active link from $g$ to $as_{i}$. There is a successor link from service $as_i$ to service $as_j$ for every WM unit such that $w_{i} \in as_{i}^{add} \cap as_{j}^{pre}$. A predecessor link from $as_j$ to $as_i$ exists for every successor link from $as_i$ to $as_j$. There is a conflicter link from $as_i$ to $as_j$ for every WM unit such that $w_{i} \in as_{j}^{del} \cap as_{i}^{pre}$. % (see Figure~\ref{fig_attention}).
Additionally, the model defines five global parameters that can be used to tune the global behavior of the network: $\pi$ is the mean level of activation, $\theta$ is the threshold for becoming active, $\phi$ is the amount of activation energy a WM unit injects into the network, $\gamma$ is the amount of energy a goal injects into the network, and $\delta$ is the amount of activation energy a protected goal takes away from the network. These global parameters make it possible to mediate smoothly between service selection criteria, such as trading off goal-orientedness for situation-orientedness, adaptivity for inertia, and deliberation for reactivity (see Listing 1.1).

\begin{pseudo-listing}[Pseudocode for the Spreading Activation Dynamics of \middleware's Attentional Mechanism (see Section F.)]
input: a set of WM units $W$, a set of goals $G$, cognitive cycle $t$
output: selected abstract service $AS$
$A$ := set of registered abstract services //$A = \{as_1... as_n\}$
$M_j$ := nil //$\forall as \in A, j \in W$ $|$ $M_j = \sum_{i=1}^n \#(as_i^{pre} \cap j)$
$X_j$ := nil //$\forall as \in A, j \in W$ $|$ $X_j = \sum_{i=1}^n \#(as_i^{add} \cap j)$
$U_j$ := nill //$\forall as \in A, j \in W$ $|$ $U_j = \sum_{i=1}^n \#(as_i^{del} \cap j)$
for each abstract service $as_i$ in $A$ do:
  $AW_{(i, t)}$ := $\sum_j\phi \cdot (1 / M_j) \cdot(1 / \#(as_i^{pre}))$ //compute activation from current WM state ($AW$)
  $AG_{(i, t)}$ := $\sum_j\gamma \cdot (1 / X_j) \cdot (1 / \#(as_i^{add}))$ //compute activation from goals ($AG$). 
  $TG_{(i, t)}$ := $\sum_j\delta \cdot (1 / U_j) \cdot (1 / \#(as_i^{del}))$ //take activation away from achieved goals ($TG$)
  $BW_{(i, t)}$ := $\sum_j as_i^{\alpha(t-1)} \cdot (1 / X_j) \cdot (1 / \#(as_i^{add}))$ //spread activation energy backward ($BW$)
  $FW_{(i, t)}$ := $\sum_j as_i^{\alpha(t-1)} \cdot (\phi/\gamma) \cdot  (1 / X_j) \cdot (1 / \#(as_i^{add}))$) //spread activation energy forward
  $as_{(i, t)}$.act := $EW_{(i, t)} + EG_{(i, t)} - TG_{(i, t)} + BW_{(i, t)}) + FW_{(i, t)}$ //total activation for $as_i$
end for
return $AS$ := $\max_{act}(A)$
\end{pseudo-listing}
\noindent\emph{F. Procedural Memory (PM)}

PM defines a set of heuristics (in the form of productions) to: 1) discover and match concrete services based on contextual information and QoS attributes; and 2) adjust the BN parameters in order to make the global behavior be more adaptive (i.e., deliberative vs. reactive, goal-oriented vs. situation-oriented, etc.) depending on the task requirements.
%
%In concordance with a bounded rationality approach, we used heuristics rather than strict rigid rules of optimization because heuristics often lead to better decisions by focusing on one aspect of the composition problem and ignoring others while considering resource scarcity and QoS. %(e.g., limited available information, time constraints, etc.).
% 
%For instance, 
Suppose there are two concrete services associated to "get-location" abstract service (e.g., Bob's phone hosts $cs_p$ and Bob's smartwatch hosts $cs_w$), and each concrete service has 2 QoS features: accuracy and latency. The accuracy of $cs_p$ is better since it uses fused location algorithms and its GPS sensor provides more precise readings, but its latency is higher than $cs_w$, so at a given moment, a heuristic production might prefer to match $cs_w$, even if this is not as accurate as $cs_p$, just because it can deliver a faster response during time-sensitive compositions. 
Regarding tuning BN parameters, PM applies the following heuristics~\cite{Romero:2011} to keep the balance between: (1) goal-orientedness vs. situation-orientedness, $\gamma > \phi$; (2) deliberation vs. reactivity, $\phi > \gamma \wedge \phi > \theta$; (3) bias towards ongoing plan vs. adaptivity, $\phi > \pi > \gamma$; and (4) to preserve sensitivity to goal conflict, $\delta > \gamma$. The corresponding values are dynamically adapted over time and using a \emph{reinforcement learning mechanism} based on the heuristic utility. 
Utility learning for a heuristic $i$ after its $n$th usage is: $U_i(n) = U_i(n-1) + \alpha[R_i(n) - U_i(n-1)] + \epsilon$, where $\alpha$ is the learning rate (default~\cite{Anderson:2004}: 0.2), $R_i(n)$ is the effective reward value given to heuristic $i$ for its $n$th usage, $\epsilon$ is a temperature (level of randomness) that is decreased over time, i.e., $\epsilon = 1/e^{(n/k)}, k=0.35$ (determined empirically~\cite{Anderson:2004}). \newline

%% ===================== Action Selection ====================== %%
% % \vspace{-0.2cm}
% \subsection{Action Selection (AS)}
% \label{sec_action}

\vspace{-0.2cm}
\noindent\emph{G. Action Selection (AS)}

AS processes different kind of actions: (1) internal actions such as goal setting (it adds new goals to both the WM and SA modules); and (2) external actions such as triggering a device's effector/actuator, and invoking the discovery mechanism to look up a concrete service and then execute it. AS uses a scheduler mechanism to sort (by priority) and execute actions in the future.

%% ===================== Cognitive Cycle ====================== %%
% \vspace{-0.2cm}
% \subsection{Cognitive Cycle}
% \label{sec_cycle}

\vspace{0.0cm}
\noindent\emph{H. Cognitive Cycle}

In mapping to human behavior~\cite{llc:2017}, \middleware's cognitive cycles operate at roughly 50 ms, although the actions that they trigger can take significantly longer to execute. A cognitive cycle starts with sensing and usually ends with selection of an internal or external action. The cognitive cycle is conceived as an active process that allows interactions between the different components of the architecture. %Thus, cognitive cycles are always ongoing. 
Deliberation, reasoning, and generation of plans in \middleware take place over multiple cascading cognitive cycles in the current situation (i.e., multiple overlapping cycles iterating at an asynchronous rate, see Listing 1.2).

\begin{pseudo-listing}[Pseudocode for COPERNIC's Cognitive Cycle (see Figure~\ref{fig_architecture})]
input: set of sensory inputs $SI$, set of user goals $G$
output: selected concrete service $CS$
$P$ := nil //set of salient percepts ($P=\{p_1..p_n\}$)
$W$ := nil //set of active units in the WM ($W=\{w_1..w_n\}$)
$D$ := nil //set of retrieved declarative units ($D=\{d_1..d_n\}$)
$A$ := set of registered abstract services ($A=\{as_1..as_n\}$)
$C$ := set of registered concrete services ($C=\{cs_1..cs_n\}$)
$R$ := nil //set of resulting actions
while remaining goals $G > 0$ do: //cognitive cycle $i$
  $P_i$ := detect, classify and temporarily store $SI$ and $G$
  $W_i$ := add salient percepts to WM ($W_i = W_{(i-1)} \cup P_i$)
  $D_i$ := cue and retrieve declarative memories using the content of WM
  $W_i$ := add declarative units to WM ($W_i = W_i \cup D_i$)
  $W_i$ := decay and filter WM units ($W_i = W_d$, where $W_d \subseteq W_i$)
  $A_i$ := focus attention and do service matching ($A_i = W_i \cap A^{pre}$)
  $C_i$ := apply heuristics (PM) and select concrete service candidates ($C{i} = A_i \cap QoS$)
  $CS$ := select a concrete service
  $R_i$ := wrap $CS$ execution as an action and add it to the set of actions
  $a$ := prioritize actions $R_i$ and pick the most relevant
  if $a$ is of type ``service-execution'' then execute concrete service's action $a.CS$
  else execute internal action
end while
\end{pseudo-listing}

% %% ===================== Context Management ====================== %%
% \subsection{Context Acquisition}

% For the purpose of service composition, this paper only is concerned about contextual raw data acquisition (i.e., we do not cover other areas of context-awareness such as context fusion, modeling, inconsistency detection, etc.). \middleware collects three different kind of contexts: physical contexts, such as accelerometer, temperature, battery, etc.; computing contexts, such as network condition, CPU, memory, etc.; and user contexts, such as location, activity, status, social relationships, etc. We used AWARE framework~\cite{Aware:2014} for acquiring physical and user contexts; and operating system primitives such as Iostat, Vmstat and Netstat for computing context.

%% ===================== Session Management ====================== %%
% \subsection{Session Management}

\vspace{-0.1cm}
\noindent\emph{I. Session Management}

\begin{figure}[h]
\vspace{-0.5cm}
\centering
\includegraphics[width=3.5in] 
{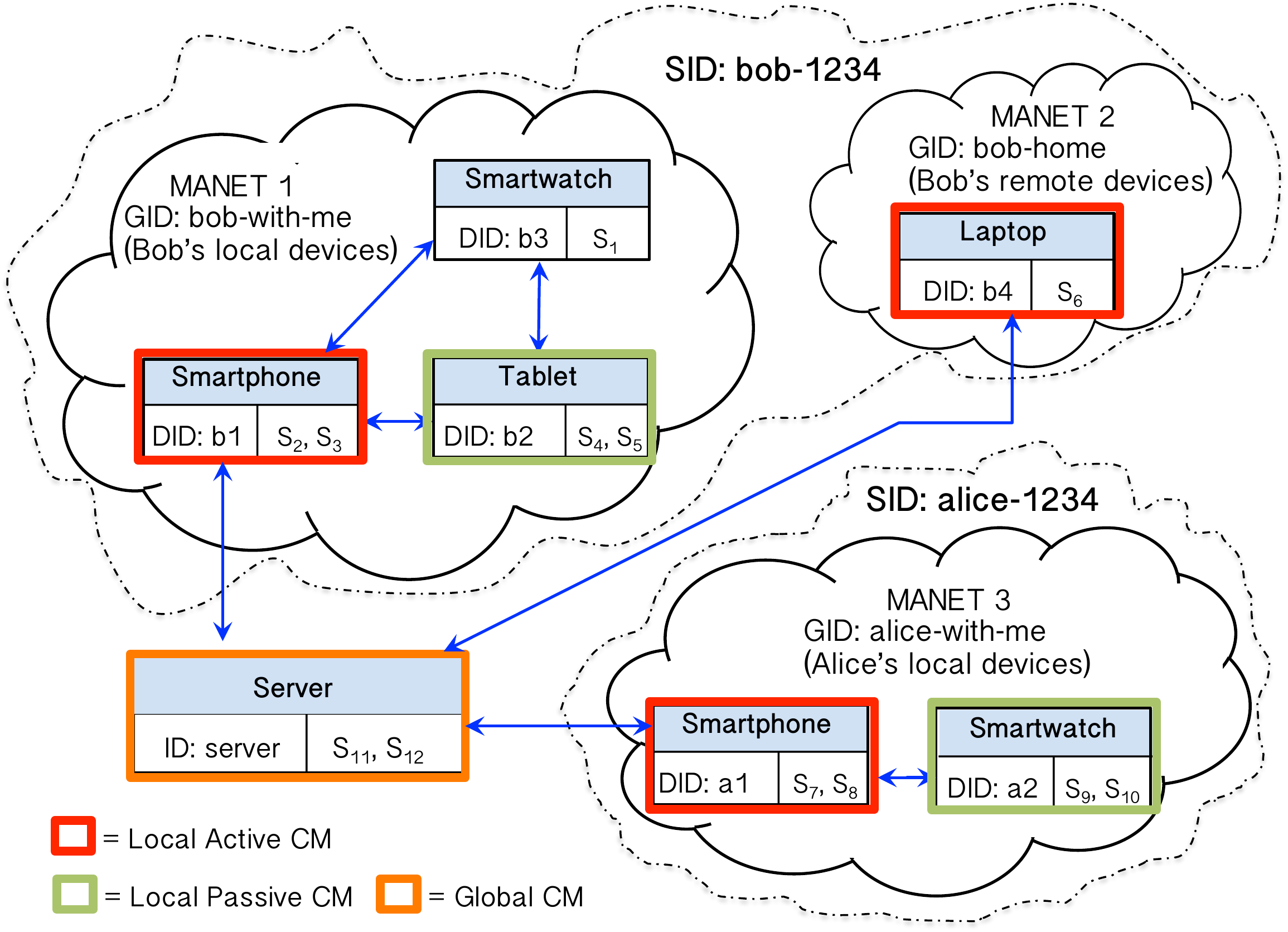}
\vspace{-0.3cm}
\caption{Infrastructure-less MPC Environment. \middleware agents are distributed across different local networks (MANET) and collaborate during composition. DID: Device ID, GID: Group ID, SID: Session ID, $S_1..S_n$: Services.}
\vspace{-0.5cm}
\label{fig_node_dist}
\end{figure}

This module manages shared sessions across multiple user's \emph{Devices} (service providers). A running instance of \middleware is hosted by each device. Using proximity discovery protocols, devices are grouped by physical nearness into \emph{Groups}. For instance, on Figure~\ref{fig_node_dist}, Group \emph{``bob-with-me"} represents the devices that Bob carries with him whereas \emph{``bob-home''} represents a Group of devices at Bob's home. 
Groups belonging to the same user are logically grouped into \emph{Sessions}, where each Session guarantees that multiple ubiquitous devices can share information and collaborate in a distributed, inexpensive and robust fashion.

%% ===================== Service Management ====================== %%
% \subsection{Service Management}
% \label{sec_service_management}

\vspace{-0.1cm}
\noindent\emph{J. Service Management}

Service management (i.e., service discovery, service coordination, and management of the information flow between service components) is performed by a \emph{Composition Manager} (CM). Service management tasks are distributed on an as-needed basis among multiple \emph{Composition Managers} (CMs) in both infrastructure-based and infrastructure-less networks. 
Unlike traditional approaches, where transferring the control from one CM to another does not consider fatal failures that prevent such transference, our approach is based on a Binary Star pattern that serves for primary-backup node failover.
That is, \middleware chooses two CMs per Group, one works in active mode while the other one in passive mode. Both active and passive CMs monitor each other and if the active CM disappears from the network, then the passive one will take over as active CM and send a request for another device to join as a passive CM.
Using a ranking function based on four levels of device capabilities, as show in Table~\ref{tab:levels}, \middleware is able to determine which devices will be the active and passive CMs, that is: $R_{cm} = Pri_{cm} \times Per_{cm}$, where $R$ is the ranking of CM $cm$, $Pri$ is a level-based priority function (i.e., Level-0 = 0... Level-3 = 3), and $Per$ is a CM performance function based on composition completeness, composition time, etc.

\begin{table}
\vspace{-0.6cm}
\caption{Composition Manager Levels}
\vspace{-0.5cm}
\centering
\begin{center}
    % \begin{tabular}{| p{2.35cm} | l | p{1.2cm} | p{1.2cm} | p{1.18cm} |}
    \resizebox{10cm}{!}{%
    \begin{tabular}{l l l l l}
    \hline
    \textbf{Feature} & \textbf{Level-0  } & \textbf{Level-1  } & \textbf{Level-2}  & \textbf{Level-3}\\
    \hline
    COPERNIC Version		& Minimal 	& Lightweight & Full 		& Full\\
    Resources (CPU, memory)	& Scarse 	& Limited 	  & Rich 		& Rich\\
    Can act as CM?			& NO 		& YES 		& YES 		& YES\\
    Cross-session composition? & NO 		& NO 	& NO  & YES\\
    Example devices				& Sensors	& Wearables   & Smartphone, Laptop	& PC, server\\
    \hline
    \end{tabular}
    }
    \vspace{-1cm}
\label{tab:levels}
\end{center}
\end{table}

\begin{figure}[h]
\vspace{-0.5cm}
\includegraphics[width=\columnwidth] 
{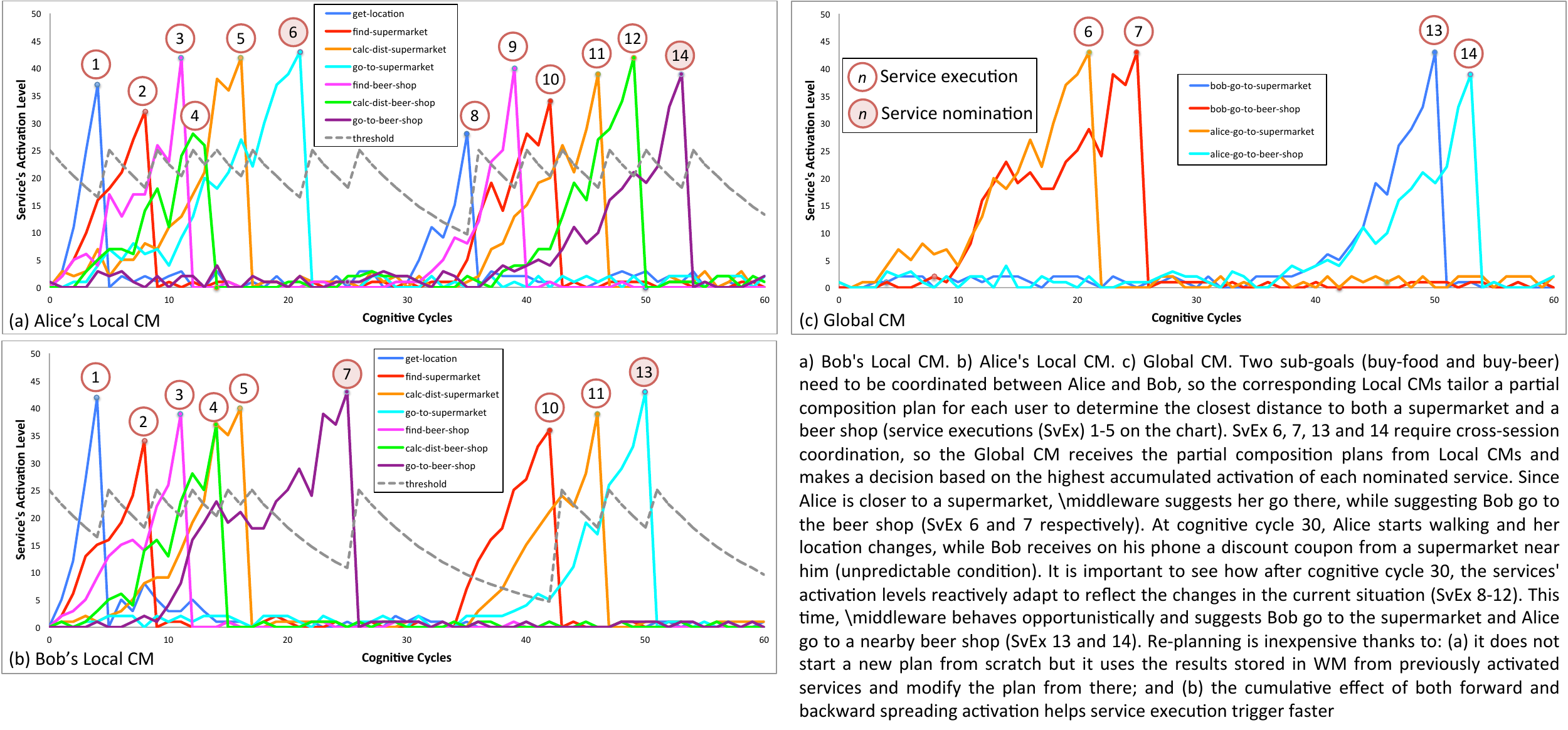}
\vspace{-0.9cm}
\caption{Spreading Activation Dynamics for the example described in Section~\ref{sec_intro}}
\label{fig_activation}
\vspace{-0.8cm}
\end{figure}

It is worth noting that only Level-3 can perform \emph{global cross-session compositions} (i.e., involving multiple users and sessions), though Level-2 and Level-1 can still perform \emph{local compositions} (i.e., within a group) so the whole composition task is distributed among multiple CMs, as shown in Fig.~\ref{fig_node_dist}. Each Local CM tailors a partial composition plan that satisfies the needs of its user by using local resources and services; then a Global CM, which can see the whole picture, receives as inputs partial plans from Local CMs and makes ``good enough'' decisions using the available information. By good enough we mean that, rather than trying to look for an optimized solution, CMs exploit opportunities that contribute to the ongoing goal/plan while adapting to unpredictable changing situations. Fig.~\ref{fig_activation} illustrates how CMs make good enough decisions as a result of the spreading activation dynamics coordinated by \middleware cognitive agents.

\vspace{-0.4cm}
\section{Evaluation}
\label{sec_evaluation}

% \subsection{Experimental Setup}

\vspace{-0.3cm}
\noindent\emph{A. Experimental Setup}

We implemented the dynamic composition overlay on the NS-3 simulator using the experimental settings on table~\ref{tab:settings}. We compared \middleware against two state-of-the-art decentralized service composition models: GoCoMo, a goal-driven service model based on a decentralized heuristic backward-chaining planning algorithm~\cite{Chen2018}; and CoopC, a decentralized cooperative discovery model that employs a backward goal-driven service query and forward service construction mechanism but does not support runtime composite service adaptation~\cite{Furno:2013}. We run 2 different experiments and measured the composition efficiency using 4 different metrics: composition time (\textbf{CT} in seconds), average memory usage of all service providers involved in the composition (\textbf{MU} in Kb), a planning failure rate (\textbf{PFR}) calculated as the ratio of the number of failed planning processes to the number of all the issued requests during the simulation cycles, and the execution failure rate (\textbf{EFR}) computed as the ratio of the number of failed executions to the number of all the successful plans.

\begin{table}[h]
\vspace{-0.5cm}
\caption{Experimental Settings}
\vspace{-1cm}
\centering
\begin{center}
\resizebox{\columnwidth}{!}{%
\begin{tabular}{ l l }
\hline
\textbf{Setting}        & \textbf{Value}                             \\
\hline
Number users (1 goal/user)   & 1, 2  \\
Service density (\# providers)   & sparse (\textbf{SD-S}): 20, medium (\textbf{SD-M}): 40, dense (\textbf{SD-D}): 60 \\
Composition Length  & 5 services (\textbf{CL-5}), 10 services (\textbf{CL-10})                                      \\
Node Mobility    & slow (\textbf{M-S}): 0-2m/s, medium (\textbf{M-M}): 2-8m/s, fast (\textbf{M-F}): 8-13m/s \\
Number of Services    & Abstract: 10, Concrete: 40 (4 per abstract service)                         \\
Communication range & 250 m                                       \\
Sem. matchmaking delay  & 0.2 (s)~\cite{Chen2018}                                       \\
Sample per experiment                   & 100 runs                                        \\
Node movement           & Random walking Monte Carlo model           \\
Pre/post-cond per service   & Random (1-4)   \\     
% Machine     & MacBook Pro 2.5GHz Intel Core i7\\
\hline
\end{tabular}
}
\vspace{-1cm}
\label{tab:settings}
\end{center}
\end{table}
%
% \subsection{Experimental Protocol}
% \vspace{0.1cm}
% \noindent\emph{B. Experimental Protocol}
%

\noindent\emph{B. Flexibility of Service Composition}

This scenario evaluates the flexibility of \middleware, GoCoMo, and CoopC during the generation of service composites while varying node mobility, service density, and service complexity (composition length). This scenario uses the configuration presented in table~\ref{tab:settings} and only one user. % (since neither GoCoMo nor CoopC support multiple users and goals). 
The experimental results are shown in table~\ref{tab:flexibility} (blue and red cells are the best and worst measurements for each category, respectively). Overall, GoCoMo got the lowest failure rates (PFR), followed by \middleware and then by CoopC, though \middleware's composition time (CT) and memory usage (MU) were the lowest in comparison to the other two approaches. 
In particular, GoCoMo got a lower failure rate (12 - 38\%) than \middleware when the mobility was slow. This difference dropped to 7 - 13\% in fast-mobility high-density scenarios thanks to \middleware is less sensitive to mobility changes because the information about participant services is stored in the WM and gradually fades away, so when, for instance, a service provider disappears and reappears later in time, the probability that this service provider is still in the WM is high (due to its activation may have not decayed entirely), so it will be able to promptly participate again in the composition without producing significant planning failures. In comparison with CoopC, \middleware got 12 - 25\% less failures due to CoopC's does not support runtime adaptation and poorly handles mobility and density changes. Regarding composition time, \middleware tailored composite services up to 42\% faster than GoCoMo and up to 71\% faster than CoopC; and it used up to 72\% less memory than GoCoMo and up to 84\% less memory than CoopC. The reason for this significant reduction in composition time and resource consumption is that \middleware is continuously filtering out the stream of incoming information (sensory stimuli), which keeps it into reasonable margins of resources usage, despite of the dynamism of the environment. 
It is worth noting that \middleware did not show a significant difference in memory usage when using a composition length of either 5 or 10 services (-4\% - 11\%) in comparison with GoCoMo  (60\% - 190\%) and CoopC (157\% - 201\%), which suggests that our approach could be smoothly scaled up.

\begin{table}[htb]
\centering
\caption{Flexibility of Service Composition}
\vspace{-0.2cm}
\label{tab:flexibility}
\resizebox{10cm}{!}{%
\begin{tabular}{l|l|l|r|r|r|r|r|r|r|r|r} 
\hline
\multicolumn{3}{l|}{}                                                                                                                                                                & \multicolumn{3}{c|}{\textbf{M-S} }                                                                                                                                        & \multicolumn{3}{c|}{\begin{tabular}[c]{@{}c@{}}\textbf{M-M}\\ \end{tabular}}                                                                                              & \multicolumn{3}{c}{\textbf{M-F} }                                                                                                                                          \\ 
\cline{4-12}
\multicolumn{1}{l}{}                                                        & \multicolumn{1}{l}{} &                                                                                 & \multicolumn{1}{c|}{\textbf{SD-S} }                     & \multicolumn{1}{c|}{\textbf{SD-M} }                    & \textbf{SD-D}                                          & \multicolumn{1}{c|}{\textbf{SD-S} }                     & \multicolumn{1}{c|}{\textbf{SD-M} }                    & \textbf{SD-D}                                          & \multicolumn{1}{c|}{\textbf{SD-S} }                     & \multicolumn{1}{c|}{\textbf{SD-M} }                    & \textbf{SD-D}                                           \\ 
\hline
\multirow{2}{*}{\textbf{COPERNIC} }                                         & \textbf{CL-5}        & \begin{tabular}[c]{@{}l@{}}\textbf{PFR (\%)}\\\textbf{CT (sec)}\\\textbf{MU (Kb)} \end{tabular} & \begin{tabular}[c]{@{}l@{}}18.2\\0.9\\\cb{63} \end{tabular}  & \begin{tabular}[c]{@{}l@{}}3.7\\\cb{0.5}\\81 \end{tabular}  & \begin{tabular}[c]{@{}l@{}}1.1\\0.8\\93 \end{tabular}  & \begin{tabular}[c]{@{}l@{}}17.5\\1.1\\67 \end{tabular}  & \begin{tabular}[c]{@{}l@{}}1.4\\1.2\\86 \end{tabular}  & \begin{tabular}[c]{@{}l@{}}1.4\\1.2\\93 \end{tabular}  & \begin{tabular}[c]{@{}l@{}}21.1\\1.1\\73 \end{tabular}  & \begin{tabular}[c]{@{}l@{}}3.3\\1.4\\88 \end{tabular}  & \begin{tabular}[c]{@{}l@{}}1.1\\1.4\\98 \end{tabular}   \\ 
\cline{2-12}
                                                                            & \textbf{CL-10}       & \begin{tabular}[c]{@{}l@{}}\textbf{PFR (\%)}\\\textbf{CT (sec)}\\\textbf{MU (Kb)} \end{tabular} & \begin{tabular}[c]{@{}l@{}}17.8\\1.2\\70 \end{tabular}  & \begin{tabular}[c]{@{}l@{}}3.7\\0.6\\86 \end{tabular}  & \begin{tabular}[c]{@{}l@{}}1.1\\0.8\\92 \end{tabular}  & \begin{tabular}[c]{@{}l@{}}17.7\\1.2\\70 \end{tabular}  & \begin{tabular}[c]{@{}l@{}}1.5\\1.2\\73 \end{tabular}  & \begin{tabular}[c]{@{}l@{}}0.5\\1.1\\85 \end{tabular}  & \begin{tabular}[c]{@{}l@{}}19.7\\1.2\\78 \end{tabular}  & \begin{tabular}[c]{@{}l@{}}3.8\\1.3\\89 \end{tabular}  & \begin{tabular}[c]{@{}l@{}}1.4\\1.9\\94 \end{tabular}   \\ 
\hline
\multirow{2}{*}{\begin{tabular}[c]{@{}c@{}}\textbf{GoCoMo}\\ \end{tabular}} & \textbf{CL-5}        & \begin{tabular}[c]{@{}l@{}}\textbf{PFR (\%)}\\\textbf{CT (sec)}\\\textbf{MU (Kb)} \end{tabular} & \begin{tabular}[c]{@{}l@{}}13.1\\1.3\\79 \end{tabular}  & \begin{tabular}[c]{@{}l@{}}3.3\\0.7\\93 \end{tabular}  & \begin{tabular}[c]{@{}l@{}}0.6\\0.9\\112 \end{tabular} & \begin{tabular}[c]{@{}l@{}}16.1\\1.3\\78 \end{tabular}  & \begin{tabular}[c]{@{}l@{}}1.2\\1.4\\93 \end{tabular}  & \begin{tabular}[c]{@{}l@{}}\cb{0.3}\\1.4\\110 \end{tabular} & \begin{tabular}[c]{@{}l@{}}18.0\\1.3\\80 \end{tabular}  & \begin{tabular}[c]{@{}l@{}}3.1\\1.3\\94 \end{tabular}  & \begin{tabular}[c]{@{}l@{}}0.9\\1.4\\114 \end{tabular}  \\ 
\cline{2-12}
                                                                            & \textbf{CL-10}       & \begin{tabular}[c]{@{}l@{}}\textbf{PFR (\%)}\\\textbf{CT (sec)}\\\textbf{MU (Kb)} \end{tabular} & \begin{tabular}[c]{@{}l@{}}16.2\\2.1\\213 \end{tabular} & \begin{tabular}[c]{@{}l@{}}2.3\\2.2\\273 \end{tabular} & \begin{tabular}[c]{@{}l@{}}0.8\\2.2\\314 \end{tabular} & \begin{tabular}[c]{@{}l@{}}24.7\\2.2\\201 \end{tabular} & \begin{tabular}[c]{@{}l@{}}1.1\\2.3\\287 \end{tabular} & \begin{tabular}[c]{@{}l@{}}0.4\\2.3\\308 \end{tabular} & \begin{tabular}[c]{@{}l@{}}22.1\\2.3\\221 \end{tabular} & \begin{tabular}[c]{@{}l@{}}3.5\\2.4\\286 \end{tabular} & \begin{tabular}[c]{@{}l@{}}1.3\\2.4\\345 \end{tabular}  \\ 
\hline
\multirow{2}{*}{\begin{tabular}[c]{@{}c@{}}\textbf{CoopC}\\ \end{tabular}}  & \textbf{CL-5}        & \begin{tabular}[c]{@{}l@{}}\textbf{PFR (\%)}\\\textbf{CT (sec)}\\\textbf{MU (Kb)} \end{tabular} & \begin{tabular}[c]{@{}l@{}}16.2\\1.8\\114\end{tabular}  & \begin{tabular}[c]{@{}l@{}}2.4\\1.9\\245\end{tabular}  & \begin{tabular}[c]{@{}l@{}}0.8\\1.9\\367\end{tabular}  & \begin{tabular}[c]{@{}l@{}}21.9\\1.9\\121\end{tabular}  & \begin{tabular}[c]{@{}l@{}}1.3\\1.8\\239\end{tabular}  & \begin{tabular}[c]{@{}l@{}}2.3\\2.1\\353\end{tabular}  & \begin{tabular}[c]{@{}l@{}}24.5\\1.9\\117\end{tabular}  & \begin{tabular}[c]{@{}l@{}}3.7\\2.1\\275\end{tabular}  & \begin{tabular}[c]{@{}l@{}}1.2\\2.2\\359\end{tabular}   \\ 
\cline{2-12}
                                                                            & \textbf{CL-10}       & \begin{tabular}[c]{@{}l@{}}\textbf{PFR (\%)}\\\textbf{CT (sec)}\\\textbf{MU (Kb)} \end{tabular} & \begin{tabular}[c]{@{}l@{}}24.0\\4.1\\325\end{tabular}  & \begin{tabular}[c]{@{}l@{}}2.3\\4.2\\476\end{tabular}  & \begin{tabular}[c]{@{}l@{}}1.3\\4.2\\593\end{tabular}  & \begin{tabular}[c]{@{}l@{}}25.2\\4.5\\332\end{tabular}  & \begin{tabular}[c]{@{}l@{}}2.4\\4.7\\488\end{tabular}  & \begin{tabular}[c]{@{}l@{}}1.2\\4.9\\605\end{tabular}  & \begin{tabular}[c]{@{}l@{}}\crd{31.8}\\5.0\\345\end{tabular}  & \begin{tabular}[c]{@{}l@{}}4.2\\5.1\\497\end{tabular}  & \begin{tabular}[c]{@{}l@{}}1.6\\\crd{5.5}\\\crd{657}\end{tabular}   \\
\hline
\end{tabular}
}
\vspace{-0.5cm}
\end{table}

%% ------------------------------------------------------- %%%
\vspace{0.2cm}
\noindent\emph{C. Adaptability of Service Composition}

In this scenario we simulated 2 users with one goal each. Then, in the middle of the composition users switched their goals (switch point). We measured the ability (in terms of planning and executing failure rates) of both \middleware and GoCoMo to adapt to the new situation and replan a different composite service for both users while using different settings for mobility, density, and composition length. In this experiment we did not consider CoopC due to it cannot perform runtime service composition adaptation. Also, for the sake of simplicity, we only used a composition length of 5 services. Since GoCoMo lacks both multi-goal and multi-user composition processing, we had to run simultaneously 2 instances of GoCoMo with one goal each, and then switched the goals at the specific switch point. The individual measurements of both instances were added up, this makes GoCoMo more comparable against \middleware. In order to demonstrate the adaptability of \middleware, we used two different configurations for its attentional mechanism. A configuration is defined as a tuple of values corresponding to the Behavior Network's parameters  described in Section 4.E, such that $C_i = \langle\theta_i, \; \pi_i,  \; \phi_i, \; \gamma_i, \; \delta_i\rangle$. Configuration $C_1$ uses the default values for the attentional mechanism: $C_1 = \langle30, 20, 20, 20, 20\rangle$, while configuration $C_2$ uses values discovered by the utility learning mechanism (described in Section 4.F) after 100 test runs: $C_2 = \langle22, 27, 42, 23, 18\rangle$. Results are presented in table~\ref{tab:adaptability}.

\begin{table}[htb]
\centering
\vspace{-0.5cm}
\caption{Adaptability of Service Composition}
\vspace{-0.3cm}
\label{tab:adaptability}
\resizebox{10cm}{!}{%
\begin{tabular}{l|l|r|r|r|r|r|r|r|r|r} 
\hline
\multicolumn{2}{l|}{}                                                                                                                                                                      & \multicolumn{3}{c|}{\begin{tabular}[c]{@{}c@{}}\textbf{M-S}\\ \end{tabular}}                                                                                                             & \multicolumn{3}{c|}{\begin{tabular}[c]{@{}c@{}}\textbf{M-M}\\ \end{tabular}}                                                                                                             & \multicolumn{3}{c}{\textbf{M-F} }                                                                                                                                                         \\ 
\cline{3-11}
\multicolumn{1}{l}{}                                       &                                                                                                                               & \multicolumn{1}{c|}{\textbf{SD-S} }                          & \multicolumn{1}{c|}{\textbf{SD-M} }                         & \textbf{SD-D}                                               & \multicolumn{1}{c|}{\textbf{SD-S} }                          & \multicolumn{1}{c|}{\textbf{SD-M} }                         & \textbf{SD-D}                                               & \multicolumn{1}{c|}{\textbf{SD-S} }                          & \multicolumn{1}{c|}{\textbf{SD-M} }                         & \textbf{SD-D}                                                \\ 
\hline
\textbf{COPERNIC - C1}                                     & \begin{tabular}[c]{@{}l@{}}\textbf{PFR (\%)}\\\textbf{EFR (\%)}\\\textbf{CT (sec)} \\\textbf{MU (Kb)} \end{tabular}          & \begin{tabular}[c]{@{}r@{}}\crd{23.1}\\45.2\\5.8\\85\end{tabular}  & \begin{tabular}[c]{@{}r@{}}5.2\\22.6\\4.2\\93\end{tabular}  & \begin{tabular}[c]{@{}r@{}}2.1\\15.9\\3.1\\110\end{tabular}   & \begin{tabular}[c]{@{}r@{}}21.1\\69.4\\6.1\\90\end{tabular}  & \begin{tabular}[c]{@{}r@{}}4.3\\31.2\\4.4\\102\end{tabular} & \begin{tabular}[c]{@{}r@{}}1.9\\21.4\\3.7\\117\end{tabular} & \begin{tabular}[c]{@{}r@{}}21.9\\\crd{78.5}\\6.4\\95\end{tabular}  & \begin{tabular}[c]{@{}r@{}}4.8\\43.9\\4.5\\111\end{tabular} & \begin{tabular}[c]{@{}r@{}}2.0\\37.0\\3.9\\123\end{tabular}  \\ 
\hline
\textbf{COPERNIC - C2}                                     & \begin{tabular}[c]{@{}l@{}}\textbf{PFR (\%) }\\\textbf{EFR (\%)}\\\textbf{CT (sec)} \\\textbf{MU (Kb)} \end{tabular} & \begin{tabular}[c]{@{}r@{}}20.7\\38.5\\4.2\\\cb{84}\end{tabular}  & \begin{tabular}[c]{@{}r@{}}4.7\\17.3\\3.1\\93\end{tabular}  & \begin{tabular}[c]{@{}r@{}}1.6\\\cb{11.3}\\\cb{2.3}\\109\end{tabular} & \begin{tabular}[c]{@{}r@{}}18.5\\58.2\\5.5\\91\end{tabular}  & \begin{tabular}[c]{@{}r@{}}3.6\\19.3\\3.9\\103\end{tabular} & \begin{tabular}[c]{@{}r@{}}1.4\\14.9\\2.9\\115\end{tabular} & \begin{tabular}[c]{@{}r@{}}19.8\\63.8\\5.6\\96\end{tabular}  & \begin{tabular}[c]{@{}r@{}}4.2\\29.8\\4.0\\112\end{tabular} & \begin{tabular}[c]{@{}r@{}}1.7\\26.6\\3.1\\120\end{tabular}  \\ 
\hline
\begin{tabular}[c]{@{}l@{}}\textbf{GoCoMo}\\ \end{tabular} & \begin{tabular}[c]{@{}l@{}}\textbf{PFR (\%)}\\\textbf{EFR (\%)}\\\textbf{CT (sec)} \\\textbf{MU (Kb)} \end{tabular} & \begin{tabular}[c]{@{}r@{}}20.3\\43.5\\5.4\\428\end{tabular} & \begin{tabular}[c]{@{}r@{}}4.2\\18.2\\3.9\\536\end{tabular} & \begin{tabular}[c]{@{}r@{}}1.4\\13.2\\3.2\\678 \end{tabular}  & \begin{tabular}[c]{@{}r@{}}18.1\\67.4\\5.9\\544\end{tabular} & \begin{tabular}[c]{@{}r@{}}3.5\\24.2\\4.3\\623\end{tabular} & \begin{tabular}[c]{@{}r@{}}\cb{1.1}\\16.8\\3.5\\701\end{tabular} & \begin{tabular}[c]{@{}r@{}}19.1\\76.3\\\crd{6.6}\\598\end{tabular} & \begin{tabular}[c]{@{}r@{}}3.9\\35.7\\4.6\\703\end{tabular} & \begin{tabular}[c]{@{}r@{}}1.6\\32.4\\3.9\\\crd{812}\end{tabular}  \\
\hline
\end{tabular}
}
\vspace{-0.7cm}
\end{table}

Despite the fact that GoCoMo had lower planning failure rates than \middleware (12\% - 42\% and 2\% - 11\% in comparison with $C_1$ and $C_2$ respectively), the memory required by it to complete the compositions was considerable higher (up to 5.7 times higher) than both configurations of \middleware. One of the main reasons for this significant divergence in memory usage is that GoCoMo's service discovery mechanism uses a cache to store the progress of resolving split-join controls for parallel service flows, which results in resource-intensive processes creating multiple potential execution branches.
On the contrary, \middleware does not keep a record of multiple execution branches and does not store in memory different workflows for each fork of the plan; it keeps in memory only one single plan that is created on-the-fly, that is, when goals or sensory stimuli (internal signals and user requests) change then it adapts to the new situation by spreading more activation to those nodes (e.g., perception, WM, and attentional nodes) that should participate in the new plan, which becomes more attractive than the current plan. Additionally, \middleware does not replan at every time step. The ``history'' of the spreading activation also plays a role in the service selection since the activation levels are not reinitialized at every time step (instead, they are accumulated over time, so when changing the focus to another goal, services that may participate in the new goal reuse their current activation and continue accumulating activation on top of it). 
Furthermore, it is important to highlight that the cost of recomposing is significantly reduced by \middleware thanks to its distributed nature where multiple agents decompose the whole problem into smaller planning objectives. Similar to the previous experiment, one of the drawbacks of our approach is that its failure rate was higher than GoCoMo's one due to the attention mechanism could dismiss some crucial information pieces at any point affecting the final decision.
Now, comparing the results of both configurations of \middleware, we can observe that in general $C_2$ outperforms $C_1$. We can infer that if $\gamma$ (the amount of energy a goal injects into the attentional mechanism) $> \phi$ (the amount of energy that the WM units inject into the attentional mechanism) then \middleware will be more goal-oriented and less sensitive to changes in the current state (e.g., changes in mobility). On the contrary, if $\phi > \gamma$ then the system will be more sensitive to changes in the current state rather than changes in the goals. Also, if $\delta$ (the amount of activation energy a protected goal takes away from the system) is significantly greater than $\gamma$ then the system will keep a stubborn position, that is, it will try to always stick with the original plan (protected goal) and will be reluctant to refocus its attention to the new goal. On the contrary, if $\gamma$ is considerable greater than $\delta$ then the system will be continuously switching from one goal to another and never will conclude any plan. Finally, reducing the activation threshold $\theta$ may help the system make faster decisions (reactive behavior), useful during time-sensitive composition; by contrast, increasing $\theta$ will slow down the reasoning process (deliberative behavior), useful for long-term composition planning. Therefore, the utility learning mechanism has to find a proper ratio between these parameters so the performance of the system is improved. The learning mechanism found a tradeoff between the parameter and discovered that, in order to make the system sensitive to both current-state changes (e.g., mobility, perceived stimuli, etc.) and goal changes, without switching indefinitely between goals, and with the ability to undone previously reached (protected) goals in order to refocus on new goals (replanning), then $\gamma$ should be slightly greater than $\delta$ at a ratio of $\approx$ 4:3; $\phi$ should be greater than $\gamma$ at a ratio of $\approx$ 2:1; and $\phi > \pi > \gamma$ where $\phi$ is greater than $\pi$ at a ratio of 14:9. When using values beyond those ratios (as $C_1$ does), \middleware's planning and execution failure rates increased considerably in comparison with GoCoMo. 

%Another type of adaptivity that \middleware exhibits is fault tolerance. This is a consequence of the distributed nature of the system. It is possible to disable services (e.g., due to a failure) and the system still would do whatever is within its remaining capabilities (i.e., creating sub-optimal but yet valid plans using the remaining services). A robust composition system should be tolerant towards such failures and degrade gracefully with increasing service/resource unavailability. Thanks to the multiple filters that act over the incoming information on each cognitive cycle, the system only focus on the most relevant information, and the decision-making process becomes inexpensive. 

As a side note, the way CoopC (and other baseline service composition models) addresses faults in composition is for the system to restart the whole process if any service has failed during the execution, of course, this solution is unable to utilize the partial results. Unlike CoopC, \middleware neither creates a long-term plan upfront, nor constructs a search tree. Instead, plans are tailored on-the-fly through activation accumulation, so that it does not have to start from scratch when one path does not produce a solution, but smoothly moves from one composition plan to another. As a result, the computation of the composition plan is much less expensive. Creating a long-term plan in advance would require too much time (especially for a cognitive agent operating in a rapidly changing MPC environment), instead, plans are emergently created by \middleware as a result of multiple cascading cognitive cycles. The main differences between our approach and both GoCoMo and CoopC are that our approach can mediate smoothly between deliberation and reactivity by determining (through learning) a tradeoff between $\phi, \gamma$, $\delta$, $\pi$ and $\theta$; and that it can perform deliberative composition by accessing long-term intentions stored in the episodic memory.
It is worth noting that there is a multiple correlation between resource consumption, execution failure rate, and planning failure rate, so the more \middleware filters out the information required for the composition the lesser resources are required during composition, the faster the composite service will be generated and, therefore, the lower the execution failure rate will be. That is, if a composite service can be planned and replanned quickly and without requiring too many resources (as \middleware does), then the discrepancies between planning and execution will be minimized and, as a consequence, the execution failure rate will be minimized as well. However, the more the information is filtered out the higher the planning failure rate will be due to the cognitive agent may dismiss critical information pieces during planning.

%% ================== RELATED WORK ====================== %%
\vspace{-0.4cm}
\section{Related Work}
\label{sec_related_work}
\vspace{-0.2cm}

In the existing literature, there are mainly three different techniques of drafting a composition plan~\cite{[6]Raychoudhury:2013}. The first one utilizes the classical planning techniques used in AI (e.g., HTN, petri-nets, state charts, rule-based, multi-agent systems, etc.). Under this approach~\cite{Santofimia:2011,davidyuk:2011}%Bertoli:2009
, the composition of atomic services into a composite service is viewed as a planning and optimization problem. The second technique uses workflows, in which a composite service is broken down into a sequence of interactions between atomic services~\cite{benmokhtar:2007}. The third technique uses direct mappings between user requests and service descriptions without needing intermmediate representations such as ontologies~\cite{Romero:2019}.
In general, first and second approaches either rely on conditional plans and can therefore handle only a limited range of non-deterministic action outcomes, or have the queries about unknown information explicitly included in the predefined service composition procedure. These plans use to be computationally expensive, have to deal with composition length and strive to optimize the resources involved. Our service composition model is not as expensive as traditional approaches because plans are constructed emergently on-the-fly as the result of both the spreading activation dynamics defined at multiple overlays of the system, and the cognitive mechanism for filtering out and focusing on the most relevant information. 
To reduce composition and execution failures while dealing with complex user requirements, existing service composition techniques investigate flexible composition planning mechanisms such as: open service discovery approaches and dynamic service planning approaches. A graph-based service aggregation method~\cite{WangXQL:09,Al-Oqily:2011} models services in an aggregation graph based on the parameter dependence of the services. It dynamically composes services to support a task in a workflow when a direct match between the task and a single service does not exist. Such a workflow may need to be generated offline by a domain expert or a composition planning engine, which is inconvenient when a change is required at runtime. Dynamic service planning approaches such as~\cite{Oh:2008,Khakhkhar:2012}%,Hibner:2007
use classic AI-planning algorithms, such as forward-chaining and backward-chaining for dynamic composition planning, and usually employ a bi-direction planning algorithm to find a path with the smallest cost from the dependency graph. However, these approaches require central service repositories to maintain service overlays, and have no support for dynamic composition replanning for composition failures.
AI-planning algorithms like Haley~\cite{Zhao:2009}, and a fuzzy TOPSIS method~\cite{Cheng:2011} have been investigated for dynamic composition planning and have features for automatic re-planning to handle failures. However AI-planning algorithms rely on central composition engines that have not yet been applied on mobile devices. In addition, they need to re-generate a new plan for failure recovery, which is time-consuming and not suitable for dynamic environments.

%
%Behavior Networks for service composition has been previously proposed in~\cite{Metrouh:2016}, however, they used it as typical AI planner (in terms of chaining pre and postconditions) and no further research on the effect that the hyper-parameters have on the mediation between service selection criteria and spreading activation dynamics was conducted.
%
%Finally, regarding cognitive service composition, main contributions~\cite{Sandeep:2008,Ali:2005} have only limited to characterize multi-agent system properties such as reputation, trust, desire, and intention, and do not address service composition as a bounded rationality problem that can be solved from a cognitively-inspired approach. 

%% ================== CONCLUSIONS ====================== %%
\vspace{-0.4cm}
\section{Conclusions and Future Work}
\label{sec_conclusions}
\vspace{-0.3cm}

We described \middleware, an agent-based model for service composition in MPC environments. Our main contribution is the implementation of a cognitive model that efficiently and dynamically orchestrates distributed services under highly changing conditions. Our approach focuses on bounded rationality rather than optimality, allowing the system to compensate for limited resources by filtering out a continuous stream of incoming information. We tested our model against state-of-the-art service composition models while modifying mobility, service density and composition complexity features, and the promising results demonstrated that a cognitively-inspired approach may be suitable for pervasive environments where resources are scarce and the smart devices have computational and hardware constraints. Our future work will mainly focus on tightly integrating a context-awareness feature into our model so cognitive agents can make more accurate decisions during  service selection. %Also, we are planning to perform exhaustive experimentation on system scalability.

\vspace{-0.5cm}

\bibliographystyle{splncs03}
\bibliography{main} 

\clearpage
% \addtocmark[2]{Author Index} % additional numbered TOC entry
% \renewcommand{\indexname}{Author Index}
% \printindex
% \clearpage
% \addtocmark[2]{Subject Index} % additional numbered TOC entry
% \markboth{Subject Index}{Subject Index}
% \renewcommand{\indexname}{Subject Index}
% \input{subjidx.ind}
\end{document}